\documentstyle[aps,preprint]{revtex}
\begin{document}
\date{Version: \today}
\title{Mesoscopic effects in the fractional quantum Hall regime}
\author{Michael R. Geller$^a$ and Daniel Loss$^b$}
\address{$^a$Department of Physics and Astronomy, University of 
Georgia, Athens, Georgia 30602-2451 \\
$^b$Department of Physics and Astronomy, University of Basel, 
Klingelbergstrasse 82, 4056 Basel, Switzerland}
\maketitle

\begin{abstract}
Edge states of the quantum Hall fluid provide an opportunity to 
study mesoscopic effects in a highly correlated electron system
that is both experimentally accessible and theoretically tractable.
In this paper we review recent work on the persistent current
and Aharonov-Bohm magnetoconductance oscillations in the 
fractional quantum Hall effect regime. 
\end{abstract}

\vskip 0.2in

\centerline{Keywords: chiral Luttinger liquid, Aharonov-Bohm effect, 
fractional quantum Hall effect}

\vskip 0.2in

\leftline{\it Corresponding author:}
\leftline{Michael R. Geller}
\leftline{Department of Physics}
\leftline{University of Georgia}
\leftline{Athens, GA 30602}
\leftline{Fax: 706-542-2492}
\leftline{Email: mgeller@hal.physast.uga.edu}

\newpage

\section{introduction}

In pioneering work, Wen\cite{Wen CS} used the Chern-Simons 
theory of the bulk fractional quantum Hall effect (FQHE) to show 
that the edge states there should be chiral Luttinger liquids (CLL).  
As in the nonchiral Luttinger liquid, electron-electron interactions 
in the CLL play an essential role and lead to physical properties 
of the FQHE edge states that can be dramatically different than that
in the integral quantum Hall effect regime. 

In this paper we present results of an ongoing study of mesoscopic
effects in the CLL. After a review of the bosonization of finite-size 
FQHE edge states, we discuss chiral persistent currents that are 
predicted to have a universal non-Fermi-liquid temperature dependence, 
and the Aharonov-Bohm (AB) magnetoconductance oscillations in the CLL.

\section{the finite-size chiral Luttinger liquid}

We begin with a brief review of the bosonization and momentum-space 
quantization of the finite-size 
CLL\cite{Haldane,Wen finite CLL,Geller and Loss}.
The dynamics of edge states in the FQHE regime is governed by the 
Euclidian action \cite{Wen review}
\begin{equation}
S_\pm = {1 \over 4 \pi g} \int_0^L \! \! dx \int_0^\beta \! \! d\tau 
\ \partial_x \phi_\pm \big( \pm i \partial_\tau \phi_\pm +
v \partial_x \phi_\pm \big),
\label{euclidian action}
\end{equation}
where $\rho_{\pm} = \pm \partial_x \phi_{\pm} / 2\pi$ is the charge 
density fluctuation for right (+) or left (--) moving electrons, 
$g=1/q$ (with $q$ odd) is the bulk filling factor, and $v$ is the 
edge-magnetoplasmon velocity. The theory (\ref{euclidian action})
can be quantized by imposing 
$[\phi_{\pm}(x) , \phi_{\pm}(x')]=\pm i\pi g \ \! {\rm sgn}(x-x').$
We then decompose $\phi_{\pm}$ into a nonzero-mode contribution
$\phi_\pm^{\rm p}$ satisfying periodic boundary conditions and a 
zero-mode part $\phi_{\pm}^0.$ The nonzero-mode part may be expanded 
in a basis of Bose annihilation and creation operators as
\begin{equation}
\phi_{\pm}^{\rm p}(x) = \sum_{k \neq 0} \theta(\pm k)
\sqrt{\textstyle{2 \pi g \over |k| L}}
\big( a_k e^{ikx} + a_k^\dagger e^{-ikx} \big) ,
\label{nonzero-mode expansion}
\end{equation}
and the zero-mode field may be written as
\begin{equation}
\phi_\pm^0(x) = \pm {2 \pi \over L} N_\pm x - g \ \! \chi_\pm ,
\label{zero-mode expansion}
\end{equation}
where $\chi_\pm$ is an Hermitian phase operator conjugate 
to $N_{\pm}$ satisfying $[\chi_\pm, N_\pm] = i.$ Here $ N_\pm \equiv 
\int_0^L dx \ \rho_\pm$ is the charge of an excited state relative to 
the ground state. Eqns.~ (\ref{nonzero-mode expansion}) and 
(\ref{zero-mode expansion}) may be used to write the normal-ordered
CLL Hamiltonian as
\begin{equation}
H_\pm = {\pi v \over g L} N_\pm^2 + \sum_{k} \theta(\pm k) v |k|
a_k^\dagger a_k .
\label{hamiltonian}
\end{equation}

The electron field operators can be bosonized as
\begin{equation}
\psi_\pm(x) = {1\over \sqrt{2 \pi a}} \,
e^{i \phi_\pm(x) /g} e^{ \pm i \pi x/ gL}, 
\end{equation}
where $a$ is a microscopic cutoff length.
The phase factor $e^{\pm i \pi x/gL}$, which has the effect of 
disentangeling the charge and phase operators in the zero-mode, is 
necessary for bosonizaton in a finite-size system; see 
Ref.~\cite{Geller and Loss} for further discussion. (Similar factors 
are known to occur in the bosonization of nonchiral finite-size 
systems \cite{Loss,Maslov etal}.) Imposing periodic boundary conditions 
on the electron field operators leads to the requirement that the
allowed eigenvalues of $N_{\pm}$ are given by \cite{Geller and Loss}
\begin{equation}
N_\pm = n g ,
\label{fractional charge}
\end{equation}
where $n$ is any integer, which means that there exists
fractionally charged excitations, as expected in a FQHE system.

\section{persistent current in the chiral Luttinger liquid}

In a macroscopic FQHE edge state, an equilibrium edge current exists \
even in the absence of an AB flux or twisted boundary conditions. The 
magnitude of this current is universal and in the absence of disorder 
is given by \cite{Geller and Vignale}
$ I_{\rm edge} = g e \omega_{\rm c} / 4 \pi 
+ e {\tilde \epsilon}_{\rm qh} / 2 \pi,$
where $\omega_{\rm c}$ is the cyclotron frequency and ${\tilde 
\epsilon}_{\rm qh}$ is the proper quasihole energy 
\cite{proper energy} of the Laughlin state at filling factor $g = 1/q$.

We now couple the edge state to an AB flux $\Phi$. The grand-canonical 
partition function of the mesoscopic edge state factorizes into a 
zero-mode contribution,
\begin{equation}
Z^0 = \sum_{n=-\infty}^\infty e^{- g \pi^2 (T_0/T)
(n-\varphi)^2},
\label{zero-mode partition function}
\end{equation}
which depends on $\Phi$, and a flux-independent contribution
$Z^{\rm p}$ from the nonzero-modes.
Here $T_0 \equiv v/\pi L$ and $\varphi \equiv \Phi / \Phi_0$,
with $\Phi_0 \equiv hc/e$ the flux quantum.
Note that if $N_\pm$ were restricted to be an integer
then the period of these equilibrium AB oscillations would be
$\Phi_0/g$.
The allowed fractionally charged excitations
(\ref{fractional charge}) are therefore responsible for
restoring the AB period to $\Phi_0$, as is well-known in
other contexts \cite{AB period}.

The edge current induced from the additional flux $\Phi$ is found
to be \cite{sante fe}
\begin{equation}
I \equiv - {\partial \Omega \over \partial \Phi}
= {2 \pi  T \over \Phi_0 } \sum_{n=1}^\infty
(-1)^n { \sin(2 \pi n \varphi) \over
\sinh(n q T / T_0)},
\label{persistent current}
\end{equation}
where $\Omega$ is the grand-canonical potential. At zero temperature, 
this {\it chiral persistent current} has an amplitude $g {ev \over L}.$
Note that $I$ is renormalized by interactions in precisely the
same way as in a nonchiral Luttinger liquid \cite{Loss}.
For $T \gg T_0$ the amplitude decays as
$ g {e v \over L} e^{-q T/T_0}.$
Because these persistent currents are chiral, there is no
backscattering from impurities and hence no amplitude
reduction from weak disorder.
The temperature dependence of the orbital magnetic response
of a FQHE edge state may therefore be another ideal system to observe
non-Fermi-liquid mesoscopic behavior.

We note here that the persistent current in a FQHE annulus (the
sum of currents from the inner and outer edge states) has been studied 
recently by Kettemann \cite{Kettemann}.

\section{Aharonov-Bohm effect in the chiral Lutinger liquid}

The first experimental observation of a CLL was made by Milliken, 
Umbach, and Webb \cite{Milliken etal}, who measured the tunneling 
current between two filling factor $1/3$ edge states in a 
quantum-point-contact geometry. As the gate voltage was varied, 
resonance peaks in the conductance were observed to have the correct 
CLL temperature dependence as predicted by Moon and coworkers 
\cite{Moon etal}, following earlier related work by Kane and Fisher
\cite{KF LL}. However, recent experiments by Franklin {\it et al.}
\cite{Franklin etal} and by Maasilta and Goldman
\cite{Maasilta and Goldman} on tunneling between FQHE edge states 
through an additional edge state circling an antidot have reported 
Fermi liquid behavior.

We have studied this problem using CLL theory and found 
that the transport properties of the quantum-point-contact system 
and the antidot system differ in two important ways: The first 
is that mesoscopic effects are very important in the latter. 
When the thermal length $L_{\rm T} \equiv v/ T$ 
becomes smaller than the circumference $L$ of the antidot edge 
state, the AB oscillations become washed out, and, at the same 
time, acquire a temperature dependence that is similar to a 
chiral Fermi liquid. An experiment performed 
at a temperature significantly above the point of crossover, 
$T_0 \equiv v/\pi L,$ is therefore {\it expected} to observe 
nearly Fermi liquid behavior for many mesoscopic quantities. 
The second difference is that in contrast with the 
quantum-point-contact geometry, where it is reasonable to assume 
that there exist conditions of destructive interference 
that lead to perfect resonances, the resonances in the antidot 
geometry, which are controlled by the AB effect, are never perfect, 
even at zero temperature. 

A scaling analysis of the AB effect in the CLL has been given in
Ref.~\cite{Geller and Loss}, following similar work by Moon 
{\it et al.} \cite{Moon etal}. The scaling behavior in 
the antidot system differs from that in the point contact.
Although the renormalization group (RG) equations 
are the same, their physical implications are different. As mentioned
above, the AB effect in a chiral system does not lead to destructive 
interference, so it is not possible to tune the bare quasiparticle 
tunneling amplitude all the way to zero. 
This means that at low enough temperatures the system
will always be in the strong-antidot-coupling regime, for all values
of the AB flux and gate voltages. The sharp non-Fermi-liquid
resonance studied in the quantum-point-contact geometry, having
a width varying with temperature as $T^{1-g}$, is not 
expected in the antidot geometry at the lowest temperatures.
Furthermore, the scale $T_0$ plays an important role. Whereas in
the quantum-point-contact geometry the universal RG trajectory
implies a one-parameter universal scaling function ${\tilde G}(X)$, 
where $X$ depends on the temperature, the RG trajectory in the antidot 
problem, which we also predict to be universal for sufficiently
low $T_0$, leads instead to a two-parameter universal scaling 
function ${\tilde G}(X,Y)$, where $Y$ depends on the size of the 
antidot and is temperature independent. The scaling function 
${\tilde G}(X,Y)$ contains all the mesoscopic effects associated 
with the finite-size edge state. In the $Y \rightarrow 0$
limit ${\tilde G}(X,Y)$ reduces to the one-parameter function
${\tilde G}(X)$ calculated by Moon {\it et al.} \cite{Moon etal} 
and by Fendley, Ludwig, and Saleur \cite{Fendley etal}.

The strong-antidot-coupling regime of the AB effect can be studied 
with perturbation theory \cite{Geller etal,Chamon etal}. Here we 
shall allow both electron and quasiparticle tunneling, the latter 
being important for the douple quantum-point-contact system studied 
by Chamon and coworkers \cite{Chamon etal}. The Hamiltonian is 
$H = H_0 + \delta H$, where $H_0 = H_{\rm L} + H_{\rm R}$ is 
a sum of Hamiltonians of the form (\ref{hamiltonian}) and
\begin{equation}
\delta H = v \sum_m \sum_i \Gamma_{mi} B_{mi} + {\rm h.c.}, 
\end{equation}
where 
$ B_{mi} \equiv e^{im \phi_{\rm L}(x_i)} e^{-im \phi_{\rm R}(x_i)}
/ 2 \pi a$
is an $m$-quasiparticle tunneling operator (for macroscopic edge
states) acting at point $x_i$.
The current $I \equiv \langle {\dot N}_{\rm R} \rangle$
to first order in $\delta H$ is found to be 
\begin{equation}
I = \sum_m 2mg \ {\rm Im} \ \! \sum_{ij} \Gamma_{mi} 
\Gamma_{mj}^* \ \!  X_{ij}^m(\omega) \big|_{\omega = mgV},
\label{IV relation}
\end{equation}
where 
$X_{ij}^m(\omega)$ is the Fourier transform of $ X_{ij}^m(t) \equiv -i 
\Theta(t) \langle [B_{mi}(t), B_{mj}^\dagger(0)] \rangle.$
Each term in (\ref{IV relation}) corresponds to a process
occurring with a probability proportional to $|\Gamma_i \Gamma_j|$.
The local terms $X_{11}$ and $X_{22}$ describe independent 
tunneling at $x_1$ and $x_2$, whereas the nonlocal terms $X_{12}$ 
and $X_{21}$ describe coherent tunneling; the AB phase naturally 
couples only to the latter.
The zero-temperature voltage dependence of (\ref{IV relation})
follows immediately from the local scaling dimension 
$\Delta = m^2 g/2$ of $e^{im \phi_\pm},$ which implies that
$X^m_{ij}(t) \sim (1/t)^{2 m^2 g}$ and hence
$X^m_{ij}(\omega) \sim \omega^{2 m^2 g -1}.$ Therefore, at
zero temperature,
$I \sim V^{2 m^2 g -1}$ for low voltages, which is the same 
power-law behavior
as predicted by the RG equations \cite{RG footnote}.

Restricting ourselves now to electron ($m=q$) tunneling, the current 
can be written as $I=I_0 + I_{\rm AB} \cos[ 2\pi 
({\mu \over \Delta \epsilon} + \varphi )].$
The exact IV relation for the $g=1/3$ CLL is given in 
Ref.~\cite{Geller etal}. Here we summarize our results for general 
$q$ as a function of temperature for fixed voltage, first for 
$V \ll T_0$ and then for $V \gg T_0$.

\subsection{Low voltage regime: $V \ll T_0$}

There are three temperature regimes here. When
$T \ll V \ll T_0$, both $I_0$ and $I_{\rm AB}$
are temperature independent but have nonlinear
behavior, varying with voltage as
\begin{equation}
I \propto V^{2q-1}.
\label{nonlinear IV}
\end{equation}

When the temperature
exceeds $V$, the response becomes linear.
When $V \ll T \ll T_0$, both $G_0$ and $G_{\rm AB}$ vary with
temperature as
\begin{equation}
G \propto \bigg({T \over T_F}\bigg)^{2q-2},
\label{low-temperature G}
\end{equation}
where $T_{\rm F} \equiv v/a$ is an effective Fermi temperature.

At a temperature near $T_0$, we find that $G_{\rm AB}$ for the CLL
displays a pronounced maximum.

Increasing the temperature further we cross over into the
$V \ll T_0 \ll T$ regime where $G_0$ scales as in
(\ref{low-temperature G}), but
\begin{equation}
G_{\rm AB} \propto \bigg({T \over T_0} \bigg)
\bigg({T \over T_{\rm F}} \bigg)^{2q-2} e^{-qT/T_0}.
\label{high-temperature G}
\end{equation}
Thus $G_{\rm AB}$ exhibits a crossover from the well-known
$T^{2q-2}$ Luttinger liquid behavior to a new scaling behavior
which is much closer to a chiral Fermi liquid $(q=1)$.
Careful measurements in this experimentally accessible regime should
be able to distinguish between a Fermi liquid  and this
predicted nearly Fermi-liquid temperature dependence.

\subsection{High voltage regime: $V \gg T_0$}

Again there are three temperature regimes. For the lowest
temperatures $T \ll T_0 \ll V$, the response is again
temperature independent and
nonlinear. The direct term varies with voltage according to
\begin{equation}
I_0 \propto V^{2q-1},
\end{equation}
as in the lowest temperature,
low voltage regime. However, the flux-dependent part of the current
is now much more interesting, involving power-laws times Bessel
functions of the ratio $V/2 \pi T_0 = \pi V / \Delta \epsilon$.
As the temperature is increased further to
$T_0 \ll T \ll V$, we find a crossover
to an interesting high-temperature nonlinear regime.
Here $I_0 \propto V^{2q-1}$ as before, but now
\begin{equation}
I_{\rm AB} \propto \bigg({T \over T_0}\bigg)^q e^{-qT/T_0} V^{q-1}
\sin \bigg({V \over 2 \pi T_0} \bigg).
\label{high-temperature nonlinear IAB}
\end{equation}
Therefore, the nonlinear response at fixed temperature
can also be used to distinguish between Fermi liquid and
Luttinger liquid behavior, even at relatively high temperatures.

When the temperature exceeds V, the response finally
becomes linear.
When $T_0 \ll V \ll T$, $G_0$ scales as in
(\ref{low-temperature G}) whereas $G_{\rm AB}$ scales
as in (\ref{high-temperature G}).

The strong-antidot-coupling regime studied here is ideal for 
experimental investigation because the exact current-voltage relation 
is known, and the low-temperature crossover from weak to strong 
coupling \cite{Geller and Loss} does not complicate the analysis. If, 
by an appropriate choice of gate voltages, the antidot system starts 
in the strongly coupled regime, then it will stay in this regime 
throughout the relevant ranges of temperature and magnetic field.

This work has been supported by the NSERC of Canada.


\begin{references}
\bibitem{RG footnote} Here one solves the RG flow equations
of Ref. \cite{Moon etal},
finding that the renormalized $m$-quasiparticle tunneling
rate at an energy scale $\epsilon$ is $\Gamma_m^\prime 
= (\epsilon / \epsilon_{\rm F})^{m^2g-1} \Gamma_m$, where
$\epsilon_{\rm F} \equiv v/a$ is the effective Fermi energy
and $\Gamma_m$ is the bare value. Cutting off the scaling
at $V$ leads to $I = |\Gamma_m^\prime|^2 {e^2 \over h} \cdot V
\sim  V^{2 m^2 g-1}.$
\end{references}
\end{document}